\documentclass[%
reprint,
superscriptaddress,
 amsmath,amssymb,
 aps,
 pra,
]{revtex4-2}

\usepackage{graphicx}
\usepackage{dcolumn}
\usepackage{bm}
\usepackage{hyperref}
\usepackage{xcolor}
\usepackage{scalerel}
\usepackage{amsmath}
\usepackage{float}
\usepackage{gensymb}

\begin{document}

\preprint{APS/123-QED}

\title{Quantum enhanced real-time sensing of protein-gold adsorption kinetics}

\author{Mrunal Kamble} 
\email{mrunal@tamu.edu}
\affiliation{Department of Physics and Astronomy, Texas A\&M University, College Station, Texas 77843, USA}
\affiliation{\mbox{Institute for Quantum Science and Engineering, Texas A\&M University, College Station, Texas 77843, USA}}%

\author{Evan Humberd}
\affiliation{Department of Chemistry and Physics, University of Tennessee at Chattanooga, Chattanooga, TN 37403, USA}%

\author{Tian Li}%
\email{tian-li@utc.edu}
\affiliation{Department of Chemistry and Physics, University of Tennessee at Chattanooga, Chattanooga, TN 37403, USA}%
\affiliation{UTC Research Institute, University of Tennessee at Chattanooga, Chattanooga, TN 37403, USA}
\affiliation{UTC Quantum Center, University of Tennessee at Chattanooga, Chattanooga, TN 37403, USA}

\author{Girish S. Agarwal}
\email{girish.agarwal@ag.tamu.edu}
\affiliation{Department of Physics and Astronomy, Texas A\&M University, College Station, Texas 77843, USA}%
\affiliation{\mbox{Institute for Quantum Science and Engineering, Texas A\&M University, College Station, Texas 77843, USA}}
\affiliation{\mbox{Department of Biological and Agricultural Engineering,Texas A\&M University, College Station, Texas 77843, USA}}%


\date{\today}
\begin{abstract}
Analyzing the kinetics of biological processes plays a significant role in understanding fundamental cellular functions. Many physics-based technologies used to study such processes are limited by the shot noise inherent to the coherent states of light. These technologies can greatly benefit by leveraging quantum probes to improve the sensitivity of measurements in cellular biology. Surface Plasmon Resonance (SPR) technique has been used effectively to achieve label-free, real-time measurements of protein binding kinetics, which constitutes an important biological phenomenon occurring near the cell membrane. Here, we demonstrate the integration of this technique with the two-mode bright squeezed state having fewer fluctuations as compared to the coherent state to improve the sensitivity of measurement in studying a protein-gold adsorption process. We show 4dB of squeezing as we record the signal-to-noise ratio as the function of time and it is maintained throughout the kinetic process. The improved signal-to-noise ratio leads to a $60\%$ improvement in the sensitivity of measuring the observable rate constant $k_{s}$. The quantum advantage as shown in terms of squeezing is achieved despite the total absorption of $74\%$ from the source until the final detection after the sensor. Overall, we provide the most practical setup for improving the sensitivity of the time-dependent measurements involved in various biological processes at the molecular level.

\end{abstract}

\maketitle


\section{Introduction}
Quantum sensing and metrology is a promising field proving its importance by improving the sensitivity of measurements in various fields ranging from phase-shift measurements, force measurements, imaging, spectroscopy, and more. It has been demonstrated that the current biosensing technologies can greatly advantage by leveraging quantum probes such as the entangled photon sources and squeezed states of light. It is possible to even reach the Heisenberg limit which is fundamental theoretical limit provided by quantum theory to enhance the spatial and temporal resolution in the field of bioimaging \cite{he2023quantum}. One specialized sub-area of quantum sensing combines the benefits of both the worlds, quantum mechanics and the plasmonic technology  \cite{lee2021quantum,tame2013quantum,giovannetti2004quantum} useful in studying fundamental biochemical processes occurring at the cellular level. There exists many other technologies such as super-resolution fluorescence microscopy \cite{zhao2022sparse}that involves labeling of the specific cellular components with fluorescent dyes, Total Internal Reflection–Fluorescence Correlation Spectroscopy (TIR-FCS) \cite{thompson2007total,gavutis2006probing} to observe events near the cell surface, nanoplasmonic biosensors\cite{mayer2011localized,anker2008biosensing,soler2020nanophotonic} involving metallic nano-films or nano-structures (mainly gold), among others pushing the limits to improve the sensing capabilities for the molecular level study of the cellular processes.

However, the Surface Plasmon Resonance (SPR) has proved to be a powerful technique in medical research mainly used to analyze events at or near a metal-dielectric interface, where the metal is typically gold film or nanoparticles and the dielectric constitutes the analyte under study \cite{karlsson2004spr,masson2020portable,masson2017surface}. It has been used effectively to achieve label-free, real-time measurements of protein binding kinetics which constitutes an important dynamic process occurring near the cell membrane \cite{yang2020label,jason2006overview,douzi2017protein}. The SPR technique is based on sensing the minute changes in refractive index $\Delta n$ of the sample which in turn gives information about concentration, affinity and the kinetic parameters of the protein binding event. Understanding the dynamic processes such as protein binding is essential in learning how diseases develop at the molecular level and finding the potential treatment strategies \cite{deberardinis2012cellular}. Protein binding plays an important role in the area of drug design and delivery. Enhancing the sensitivity of the SPR instrument for analyzing protein binding kinetics can influence the efficacy and safety of a drug and can significantly advance the field of medical research \cite{wanat2020biological,schmidt2010significance,zeitlinger2011protein,phillip2012protein}.

Most of the current bio-sensing techniques including SPR systems use coherent source of light for which the sensitivity of measuring the refractive index $n$ is such that $\Delta n\propto \frac{1}{\sqrt{N}}$, where N is the total number of photons interacting with the sample. This is called the Shot Noise Limit (SNL). It has been shown that the performance of SPR technique is highly dependent on the noise characteristics of the light and that the SPR sensors are reaching their theoretical limit \cite{wang2010shot,piliarik2009surface,kolomenskii1997sensitivity}. One of the ways to improve the sensitivity of measurements is by increasing the number of photons. However, this might not be always feasible due to the common issue of photo-damage where higher intensities of light  might destroy or alter the sample under study\cite{neuman1999characterization}. An effective way to improve the signal-to-noise ratio (SNR) is by reducing the noise and this is where quantum probes prove extremely useful. Marino et al. demonstrated quantum enhancement of $56\%$ when compared with the corresponding classical configuration of state-of-the-art plasmonic sensor\cite{dowran2018quantum} using the squeezed light. Recently, it is extended to apply for an array of sensors simultaneously or measure multiple parameters independently\cite{dowran2024parallel}. A theoretical study by Mpofu et al. compares the advantage of using different quantum probes specifically for the Surface Plasmon Resonance (SPR) sensing technique \cite{mpofu2023enhanced}. A recent study by the same group utilizes the single photons generated via Spontaneous Parametric Down Conversion (SPDC) to probe the protein-gold binding kinetics using the SPR sensor\cite{lee2018quantum,mpofu2022experimental}. Thus improving the  SNR is the key to improving the sensitivity of current technology.


In our work we consider the Two Mode Bright Squeezed State (TMBSS) because of its continuous wave nature allows for a higher number of photons as compared to the discrete quantum sources while still maintaining the quantum correlations between the two modes. It can enhance the sensing capabilities specifically by reducing the fluctuations associated with the intensity difference signal. Pooser et al. has demonstrated a quantum plasmonic sensor using the Kretschmann configuration and the TMBSS source to measure the shift in SPR dip for samples with different refractive index \cite{pooser2016plasmonic}. However, no previous study has experimentally investigated the quantum advantage of TMBSS in the context of time dependent SPR sensing to measure the protein-gold adsorption kinetics. Protein adsorption on gold induces a response similar to the protein binding process as it changes the refractive index of the sample when the protein associates with or dissociates from the gold film. Any further references to protein binding to the gold film are written in the context of the adsorption process.

In this paper, we present the most practical setup to integrate the TMBSS along with the SPR sensor and give proof of principle experiment showing the quantum advantage which stays almost constant throughout the protein adsorption process. Here, we mainly focus on the association of the protein with the gold surface. This setup can be used to measure the kinetic parameters with higher sensitivity than its classical counterpart when working in the shot noise limited region. We first explain in detail the workings of the SPR sensor that detects minute changes in the refractive index of the sample in contact with the gold film. The sensitivity of measurement of refractive index is directly related to the sensitivity of measuring the kinetic parameters. We then explain the noise reduction achieved using the TMBSS in the form of squeezing. We describe the experimental setup along with the data processing performed to measure the SNR. We present our findings from plotting the SPR dip useful in deciding the incident angle on sensor to quantifying the SNR as a function of time as the Bovine Serum Albumin (BSA) protein binds to the gold film on the sensor. We show improvement in the sensitivity of estimating the observable rate constant of the binding process. Finally we discuss the challenges, the significance as well as the future implications of the setup described in this study.  

\section{Schematics of the quantum enhanced surface plasmon sensor}
\begin{figure}[b!]
    \includegraphics[width=8cm]{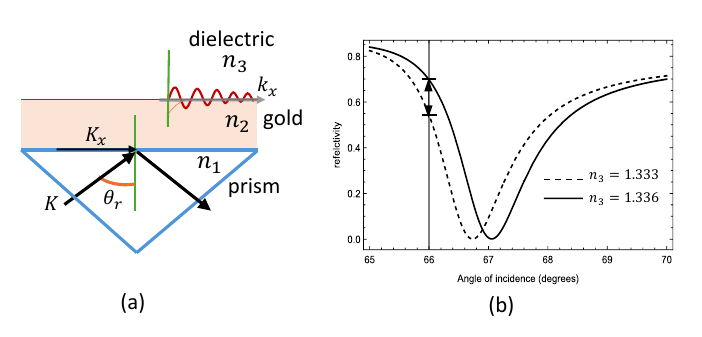}
    \caption{(a) Kretschmann geometry - The Total Internal Reflection at the prism-metal interface generates an evanescent field on the other side which couples with the charge density oscillations under the right conditions. For a given frequency of laser ($\omega$) and the given gold thickness ($h$), there is a resonant incident angle $\theta_{r}$ at which the SPPs absorb the maximum energy giving the characteristic absorption dip. This angle is sensitive to the refractive index. (b) Numerical simulation for SPR absorption dip where the reflectivity of the prism is plotted against the angle of incidence. Small increase in refractive index shifts the absorption dip towards right. }
    \label{fig:spr}
\end{figure}

Surface Plasmon Resonance is a well-known technique in the field of bio sensing. It is based on the phenomenon where the surface plasmon (electron oscillations) at the metal-dielectric interface gets coupled with the evanescent wave generated due to an incident electromagnetic field \cite{pluchery2011laboratory,maier2007plasmonics,kroo2016surface}. This results in the formation of Transverse Magnetic (TM) wave confined to the metal-dielectric interface called as the Surface Plasmon Polariton (SPP) wave. Although the phenomenon could be observed using different coupling arrangements, the most common setup used is the Kretschmann geometry as shown in Fig.~\ref{fig:spr}(a). Here the prism, the gold film, and the dielectric medium form the apparatus useful in sensing the refractive index of the dielectric. The SPP wave propagates along the x-axis and decays exponentially on both sides of the interface along the z-axis as shown in Fig.~\ref{fig:spr}(a). 
The phase matching conditions necessary for the coupling between the evanescent wave and the surface plasmon could be understood using the dispersion relation for the surface plasmon given by,
\begin{equation}\label{disp_SP}
    k_{x}=\frac{\omega}{c}\sqrt{\frac{\epsilon_{2}(\omega)\cdot\epsilon_{3}(\omega)}{\epsilon_{2}(\omega)+\epsilon_{3}(\omega)}}
\end{equation}
where, 
$\epsilon_{2}(\omega)$ is the dielectric function for the metal (here gold) and $\epsilon_{3}(\omega)$ is the complex dielectric function for the sensing medium (BSA solution). The real part for the $\epsilon_{2}$ for gold is a negative value. Considering Eq.~(\ref{disp_SP}), if the incident field is coming from air $K=\frac{\omega}{c}$ the electric field will not be able to couple with the charge density oscillations since, $k_{x}>K_{x}$. The use of the prism facilitates this coupling due to the phenomenon of Total Internal Reflection (TIR) which gives rise to the evanescent field which penetrates through the metal and couples with the surface plasmon thus giving rise to the SPPs. The dispersion relation for the incident field now is given by,
\begin{equation}\label{disp_field}
    K_{x}=\frac{n_{1}\omega}{c} \sin{\theta}
\end{equation}
where, $n_{1}$ is the refractive index of the prism. Thus the condition for the SPR is given by equating Eq.~(\ref{disp_SP}) and Eq.~(\ref{disp_field}),
\begin{equation}\label{disp_spp}
    n_{1}\sin{\theta_{r}}=\sqrt{\frac{\epsilon_{2}(\omega)\cdot\epsilon_{3}(\omega)}{\epsilon_{2}(\omega)+\epsilon_{3}(\omega)}}
\end{equation}
Thus, at a particular angle of incidence ${\theta_{r}}$ after the critical angle the SPP are formed leading to the absorption of energy from the incident field. This is called the Surface Plasmon Resonance angle. As we plot the reflectivity of this sensor against the angle of incidence, we can see an absorption dip at the angle of resonance. Thus excitation of the SPP wave is highly sensitive to the refractive index of the dielectric medium at the interface. If we keep the thickness of the metal and the frequency of the laser constant, even the slightest change in the refractive index of the sample can be detected using the shift in the resonance dip. This is shown by the plot in Fig.~\ref{fig:spr}(b) which is based on the theoretical framework called Transfer-matrix method \cite{born2013principles} which describes propagation of an electromagnetic wave through a homogeneous film situated between two semi-infinite media as explained in detail in the appendix A. Note that the reflectance curve is more linear on the left side of the resonance angle as compared to the right side. Operating in this region produces higher shift in intensity measurement for a small shift in the resonance angle and avoids saturation of the measurements. As shown by the marker in Fig.~\ref{fig:spr}(b) we operate on the left side of the resonance angle as it allows for a higher dynamic range in measurements accommodating various analyte concentrations without requiring frequent recalibration. This region offers more stability considering the mechanical vibrations or thermal drift allowing for more reproducible and reliable measurements. 
 
This technique is already used to study molecular interactions and other parameters useful for bio-sensing. It can be used to measure the affinity of a particular sample, its concentration, specificity, and the rate of interaction. Here we use this method to study the rate at which the protein adsorbs on to the gold film, which serves as a preliminary proof of concept that could be extended to study other biokinetic processes. If we lock the angle of incidence on the left side of the absorption dip, the changes in $n_{3}$ as the protein binds to the gold, could be detected by plotting the change in reflectivity as a function of time. This in turn could be translated to find the rate of the association and the dissociation of the protein as it binds and unbinds from the gold surface. 

Although an established technique, any protein binding parameter measured using a coherent source of light is shot noise limited. Even when all the technical noise is removed, the shot noise that is inherently related to the photon statistics limits the ability to distinguish two signals within a certain range limited by the fluctuations associated with the signal. The sensitivity of the estimation of parameters studied in the protein binding analysis could be improved by using quantum probes such as the TMBSS source instead of the coherent source. Next, we elaborate on the quantum advantage of using TMBSS light.

The two-mode bright squeezed state referring to the intensity difference squeezing is a well-known quantum state that has fewer fluctuations compared to the coherent state. It could be generated using different non-linear processes with atomic systems \cite{mccormick2007strong} being the most common. Here we use a rubidium atomic vapor as a nonlinear medium in which a coherent field and a vacuum field interact via the Four-Wave Mixing (FWM) process to generate two quantum correlated beams namely the probe and the conjugate. 
\begin{figure}[t!]
    \centering
    \includegraphics[width=8cm,height=6cm]{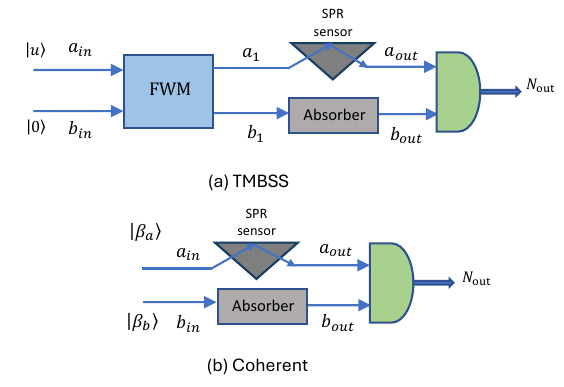}
    \caption{(a) Schematics of Surface Plasmon Resonance sensor using TMBSS light. The probe beam passing through the sensor undergoes an absorption due to SPP and the output Heisenberg operator is denoted by $a_{out}$. The conjugate beam intensity is adjusted according to the probe using an absorber and the associated Heisenberg operator is denoted by $b_{out}$. The resulting photo current is subtracted and constitutes the signal for the experiment given by $\langle{N}_{out}\rangle=\left\langle a_{out}^{\dagger}a_{out}\mathbin{-b_{out}^{\dagger}b_{out}}\right\rangle$. (b) Schematics of SPR sensor using coherent light replicating the experimental conditions for probe and conjugate. }
    \label{fig:TMBSS}
\end{figure}
As shown in Fig.~\ref{fig:TMBSS}(a), the two-mode bright squeezed state which ideally can be described as two single modes is defined using the Heisenberg picture given by the equations,
\begin{eqnarray}
 {a_{1}}&=& {a_{in}}\cosh r+ {b_{in}^{\dagger}}\sinh r, \nonumber\\
 {b_{1}}&=& {b_{in}}\cosh r+ {a_{in}^{\dagger}}\sinh r.
\end{eqnarray}
where $r$ is the squeezing parameter and $a_{1},a_{in}$ are the Heisenberg operators representing the electromagnetic field propagation as described in Fig.~\ref{fig:TMBSS}(a). This parameter $r$ is proportional to the pump power and the third order nonlinear susceptibility $\chi^{(3)}$ which describes the four wave mixing process. The pump is responsible for the FWM process. Notice that the input modes, the coherent state  $\left|u\right\rangle$ and the vacuum state $\left|0\right\rangle$ are the single modes. Using the balanced detection technique as described in \cite{kamble2024quantum}, the signal is given by $\langle{N}_{out}\rangle=\left\langle a_{out}^{\dagger}a_{out}\mathbin{-b_{out}^{\dagger}b_{out}}\right\rangle$ and the associated fluctuations are given by $\Delta N_{out}=\sqrt{\left\langle N_{out}^{2}\right\rangle -\left\langle N_{out}\right\rangle ^{2}}$.

Thus the quantum advantage here could be expressed in terms of the squeezing which is the measure of reduction in noise in dB. It is defined as,
\begin{equation}\label{Squeezing}
  S(r) = -10 \log_{10} \left(\frac{(\Delta N_{out})^2_{\scaleto{TMBSS}{4pt}}}{(\Delta N_{out})^2_{\scaleto{Coh}{4pt}}}\right) 
\end{equation}
where, $(\Delta N_{out})^2_{\scaleto{Coh}{4pt}}$ and $(\Delta N_{out})^2_{\scaleto{TMBSS}{4pt}}$ are the variance of the TMBSS and the coherent signal respectively. $(\Delta N_{out})^2_{\scaleto{Coh}{4pt}}$ is measured based on the classical sensor configuration as shown in Fig.~\ref{fig:TMBSS}(b).

To maintain the squeezing conditions before and after the sensor we match the absorption on the conjugate beam with that on the probe.
Next we explain the experimental set up in detail used to study the protein-gold adsorption process as an example of a biokinetic process using the Two Mode Bright Squeezed State as well as using the coherent state.

\section{Experimental set up}
\begin{figure*}[hbt!]
\includegraphics[width=15cm, trim=0 0 0 0cm]{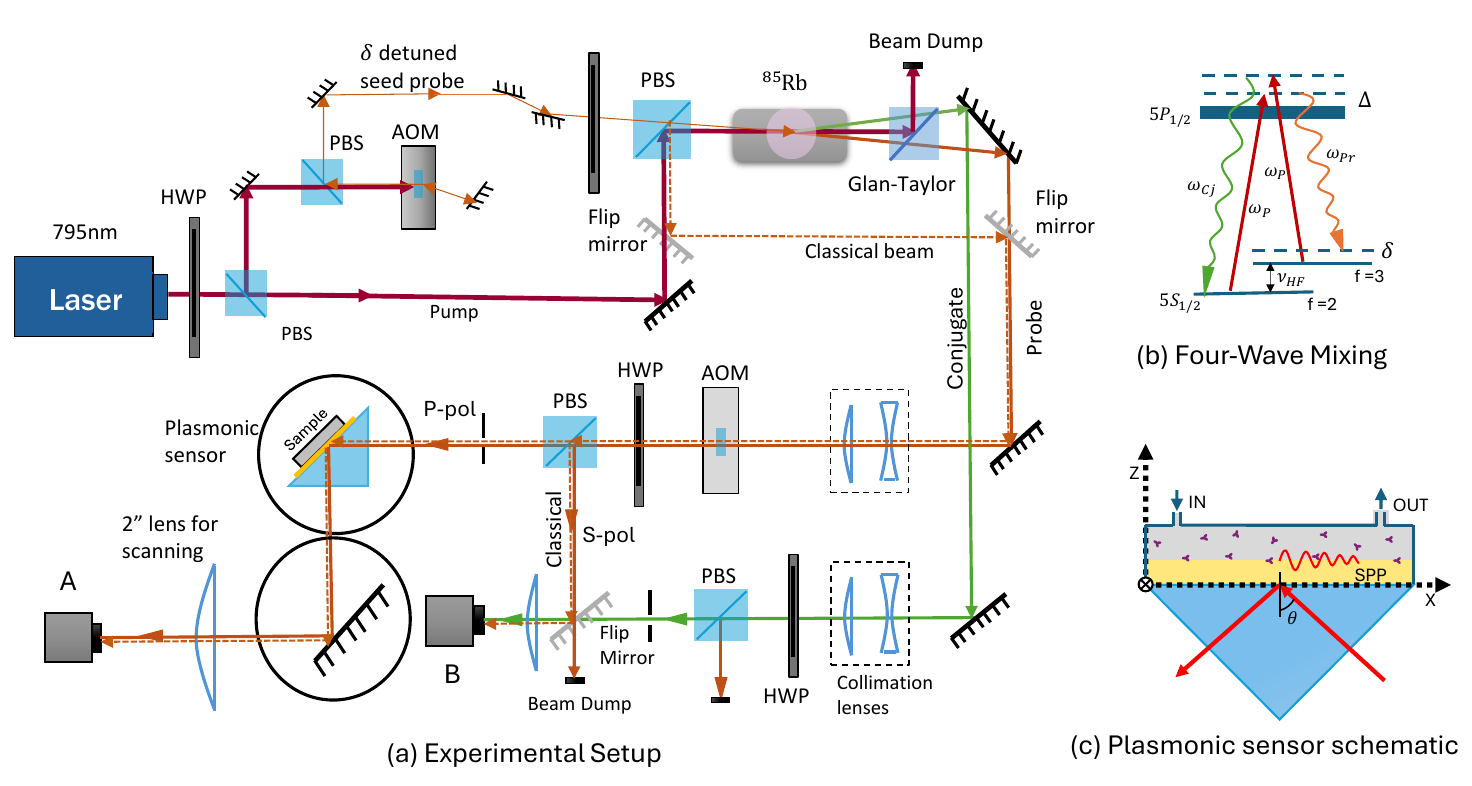}
    \caption{(a) Experimental setup showing the TMBSS generation using the Rubidium vapor and its application to study protein-gold binding process using SPR sensor. The solid lines after the Rb cell indicate the probe (orange) and the conjugate (green)beams. The intensity and the noise measurements at detector A and detector B are subtracted during the post-processing of the data. The flip mirrors indicated in gray are flipped up (active) only when performing the classical measurement. The classical beam (dotted line) splits into two beams at the Polarizing Beam Splitter (PBS) and the intensity is adjusted using the Half Wave Plate (HWP) to replicate the intensities in TMBSS case. }(b) The $\chi^{(3)}$ FWM process inside a Rb85 atom (c) Schematic of the flow cell and the sensor arrangement.
    \label{fig:exp setup}
\end{figure*}
The experimental setup we used for the quantum-enhanced protein-gold binding analysis can be divided into three parts. Firstly we generate the two mode bright squeezed state using a non-linear process occurring inside a cell containing the Rb85 atomic vapor. This process generates two beams namely the probe and the conjugate that are quantum correlated. In the next step the probe beam passes through the Surface Plasmon Resonance (SPR) sensor where the light matter interaction at the sample-sensor interface enables the study of protein binding process. In the final step, the data is collected using the balanced detection method as theoretically described in the previous section. Each step is further explained below. 
\subsection{Generating Two-Mode Bright Squeezed State}
To generate the Two Mode Bright Squeezed State we use a $\chi^{(3)}$ process called degenerate Four Wave Mixing (FWM) in which annihilation of two photons of same frequency leads to the creation of two new photons of slightly different frequencies. This non-linear process occurs inside the Rb85 atomic vapor contained in a 2.5 cm long cylindrical vapor cell. The FWM process is explained in detail in appendix B.

As shown in the upper half of  Fig.~\ref{fig:exp setup}(a) two beams namely the pump and the seed probe interact with the Rb-85 atoms generating two quantum-correlated beams called the probe and the conjugate. In this process, four control parameters - $\lambda_{pump}$ ,$\lambda_{seed}$, pump-probe alignment, and temperature of the cell are optimized to achieve the squeezing of 7.8 dB as measured at the source just after the cell. Please refer\cite{li2021quantum,li2022quantum,mccormick2007strong,mccormick2008strong} for more details of generating the TMBSS source using FWM in Rb85 atomic vapor.
The flip mirrors in Fig.~\ref{fig:exp setup}(a), create a path for the classical beam which technically is the seed probe diverted just before entering the Rb cell. The further arrangements constitute the application of the twin beams to study the protein-gold binding process. 
\subsection{Applying TMBSS to the SPR sensor}
At the output of the Rb85 cell, we have three beams namely the probe, the conjugate, and the residual pump. As shown in Fig.~\ref{fig:exp setup}(a), the residual pump is sent to the beam dump using the Glan Taylor beam-splitter which has the extinction ratio of $2\times10^5: 1$. The twin beams are then aligned and collimated to be used along with the SPR sensor to study the protein binding process.  

As shown in the lower half of Fig.~\ref{fig:exp setup}(a), the probe beam passes through a half wave plate and the  Polarizing Beam Splitter (PBS) so that only the p-polarized light is incident on the sensor. We place an Acousto-Optic Modulator (AOM) in the probe path, which modulates the intensity to fluctuate at the frequency of 2MHz. This is an important step to be able to obtain the signal as well as the noise on the same plot after data processing as explained in the detection and data processing section. 

Next, the probe beam reaches the sensor used to study the BSA protein adsorption on gold film. The sensor is assembled using commercially available SPR chip along with the on-campus facilities as explained in appendix C. The flow cell cavity as shown in Fig.~\ref{fig:exp setup}(c) is filled up with the required sample using the tubes connected via a peristaltic pump. In this paper, we use $10\%$ BSA solution which was prepared by dissolving 1gm Bovine Serum Albumin (BSA) Lyophilized powder into 10 ml of deionized water. We wait for approximately one hour to ensure complete dissolution and uniformity of the solution before injecting the fluid into the flow cell. The sensor assembly is then installed on a rotation stage such that the reflectivity can be measured at different incident angles. All the angles are measured using the incident angle of $\theta = 45 \degree$ as the reference angle which is experimentally fixed using two irises. 

The conjugate beam which is used as a reference, passes through the half-wave plate and the PBS to match the absorption in the probe path. Both the beams are focused on two separate PDA10A2 fixed gain Si detectors from Thorlabs. 
\subsection{Detection and data processing}
\begin{figure*}[hbt!]
\centering
\includegraphics[width=18cm]{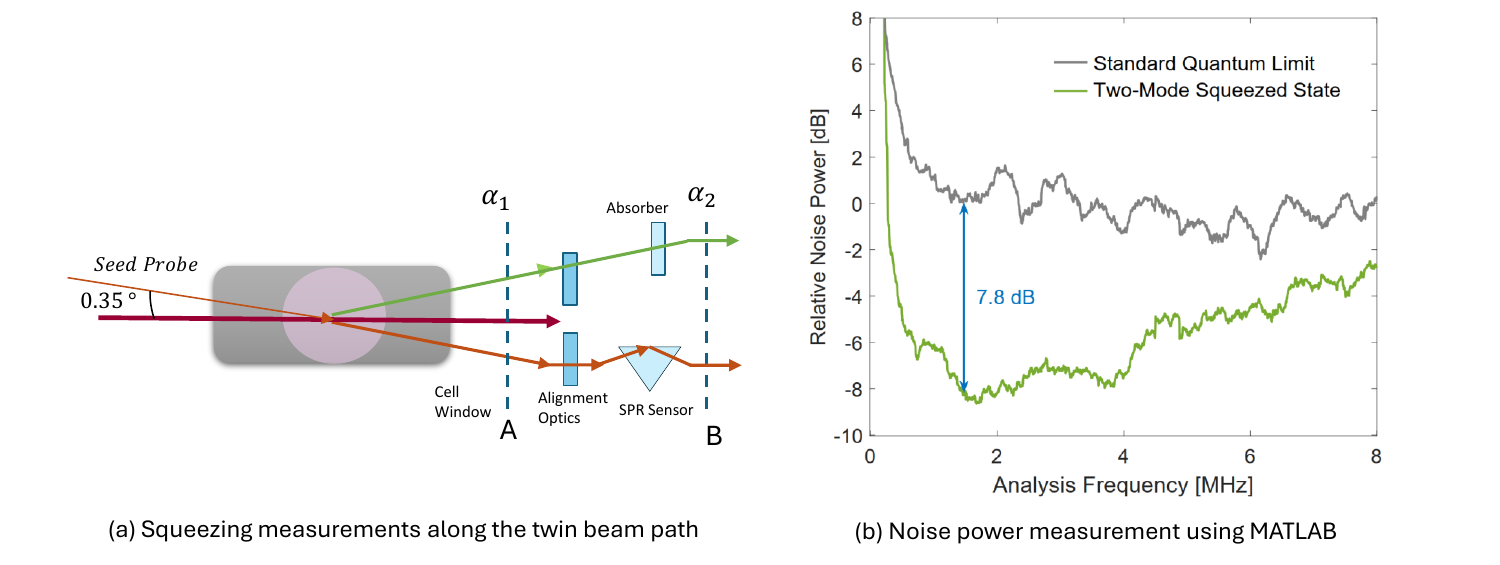}
    \caption{(a) Squeezing measurements along the propagation path of the twin beams. (b) The difference in the Standard Quantum Limit and the twin beam noise as measured at point A after post-processing the data using MATLAB.}
    \label{fig:sq measurement}
\end{figure*}
The data was collected using only the oscilloscope and was post processed with MATLAB. Each detector is connected to two channels on the oscilloscope collecting the AC voltage as well as the DC voltage data. We used the DC data mainly to see the absorption dip for water and to visualize the change in absorption during the protein binding process. The AC data which records the fluctuations in the individual signals as a function of time was converted into frequency domain using the FFT (Fast Fourier Transform) function in MATLAB. This data is used to find the twin beam noise as well as the Shot Noise Level at different points after the Rb cell. The twin beams undergo an absorption of $\alpha_{1}=0.08$ at the window of the cell. The difference between the two noise power as measured right after the cell at point A as shown in Fig.~\ref{fig:sq measurement}(b) was 7.8 dB.  Next, the twin beams need to pass through a series of optics before reaching the SPR sensor. Here it looses another $38\%$ of the intensity along with the $20\%$ loss at an off resonance angle inside the sensor giving total absorption $\alpha_{2}=0.5$. The available squeezing at the wings of the absorption dip is around 4.5 dB as measured after the sensor at point B.
\begin{figure}[hbt!]
\centering
\includegraphics[width=7cm]{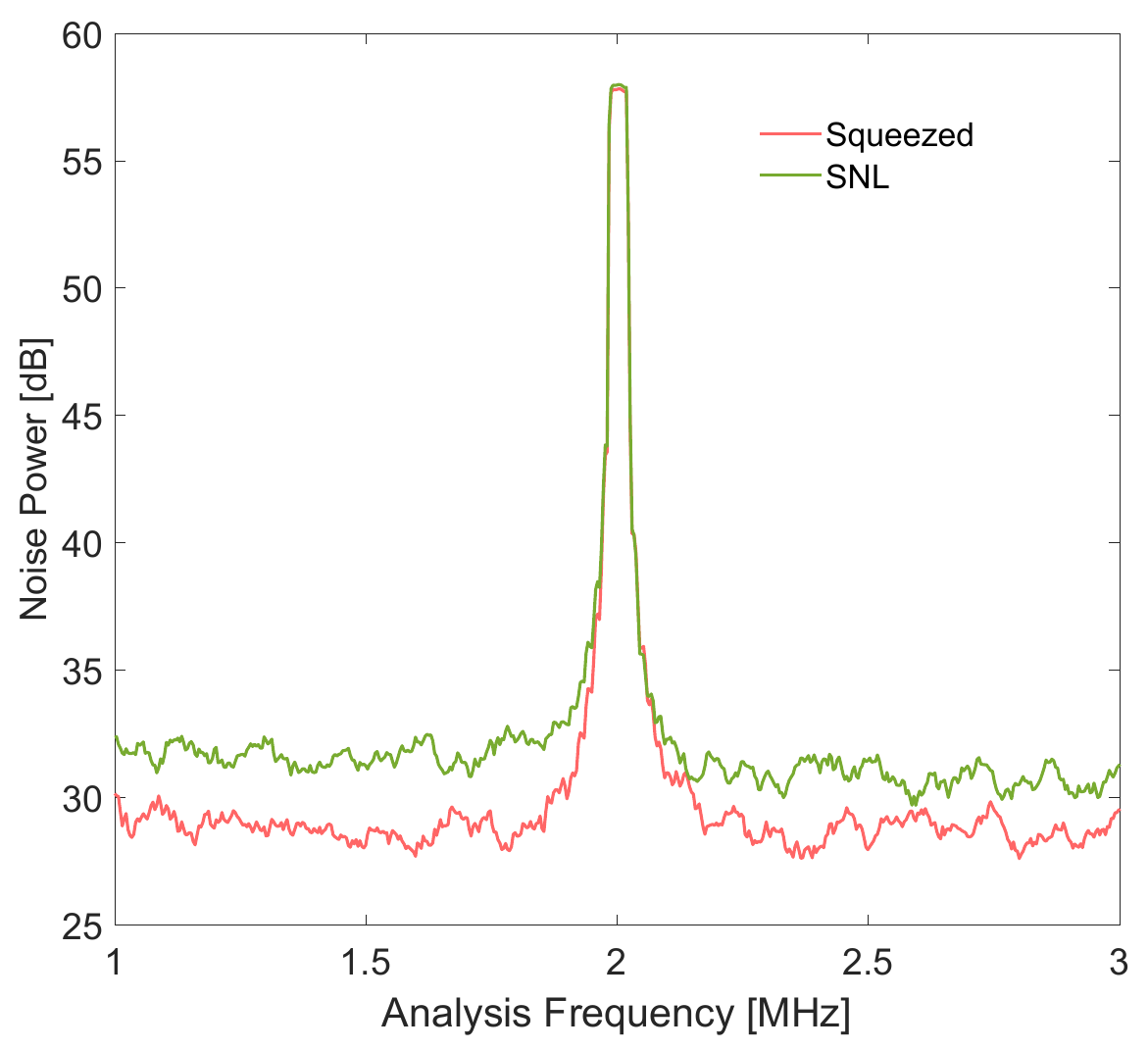}
    \caption{AOM modulation to view the sensor signal as well as noise on the same plot. }
    \label{fig:AOM}
\end{figure}
Finally, we measure the squeezing after the sensor at point B as we lock the incident angle where the reflectivity of the prism r is 0.46 $(\alpha=0.54)$. This also constitutes a starting point for the protein biding process. Thus for absorption of $\alpha_{2}=0.71$ we find the squeezing to be around 4 dB. Using the data for the noise (both coherent as well as TMBSS) and the AOM modulation as sensor signal we measure the signal to noise ratio  as a function of time. As shown in Fig.~\ref{fig:AOM} the peak amplitude is taken as the signal and the side bands correspond to the noise for a particular absorption  coefficient of the sample. It has been shown before \cite{pooser2016plasmonic} that using the amplitude of the AOM modulation as the signal which is associated to the intensity difference allows us to view the signal and the noise on the same plot plus it helps in removing the technical noise. This forms the complete setup to collect the data involved in studying the kinetics of the biological processes using the Two-Mode Bright Squeezed State along with the Surface Plasmon Resonance sensor.

 \section{Results}
\begin{figure*}[hbt!]
\centering
    \includegraphics[width=14cm]{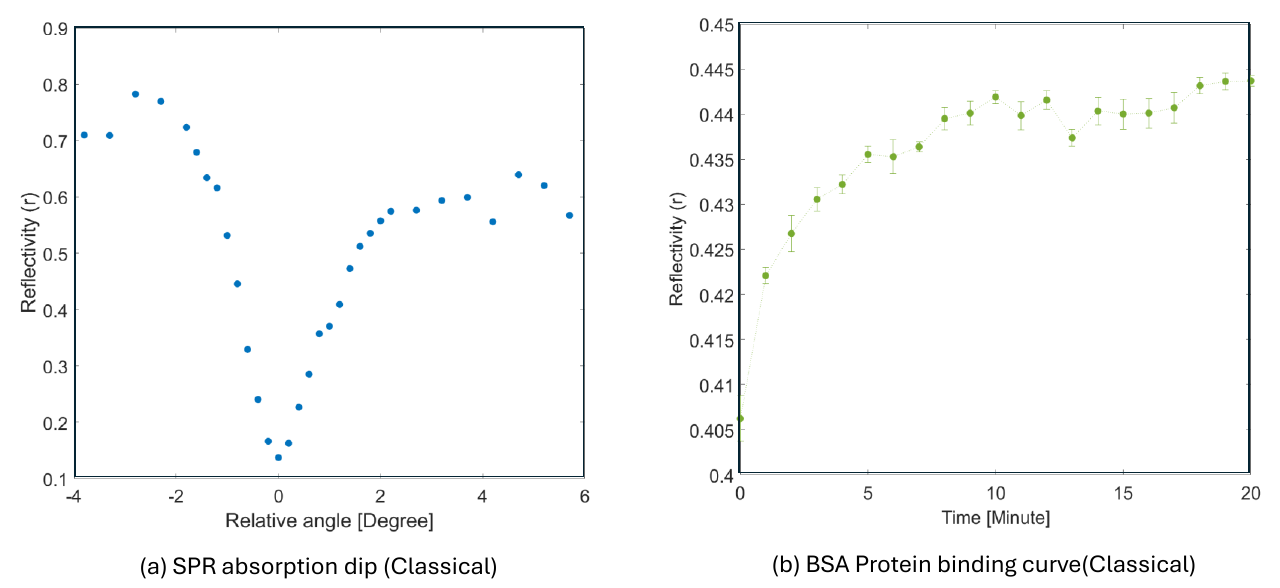}
    \caption{(a) The reflectivity of the sensor is recorded using the classical beam in the probe path. The scanning angle plotted on x-axis is measured with respect to the angle of resonance for deionized water such that zero corresponds to the incident angle $\theta=66.02 \degree$. (b) The reflectivity of the sensor is recorded against time keeping the scanning angle fixed at 0.8 degrees left of the resonance angle. The reflected intensity increases as the BSA protein binds the gold surface on the sensor. }
    \label{fig:classical}
\end{figure*}
\textit{Classical results}: Firstly, we found the Surface Plasmon Resonance (SPR) angle for deionized water (n=1.33) by plotting the reflectivity of the sensor against the scanning angle (angle made with the normal of the prism-air interface) using a coherent source. The sensor was aligned such that the classical beam in the probe path is incident at zero degrees to the normal of prism-air interface (incident angle $\theta=45 \degree$) and reflects at a right angle to the incident beam. We fix this position using two irises on the reflected path as a reference point. We recorded the reflectivity of the sensor which is the ratio of reflected intensity with the incident intensity against the scanning angle at the prism-gold interface as shown in Fig.~\ref{fig:classical}(a). We collected the data in the scanning range of $29\degree\text{--} 35\degree$ to the normal of the prism-air interface with increments of $0.2 \degree$. This gives the incident angle at the gold-prism interface in the range $\theta=63.72\degree\text{--}69.63\degree$. We find that the intensity reaches its minimum at an angle of $66.02\degree$ ($\theta_{r}$) at which point around $90\%$ of the intensity is absorbed by the excited SPPs. This aligns closely with the theoretical expectations as shown in Fig.~\ref{fig:spr}(b). The $0.7 \degree$ deviation from the theoretical simulation can be attributed to experimental errors corresponding to the initial angle. For better comparability the angle on the x-axis was converted into the normalized scale such that the angle of resonance $(\theta_r)$ is considered to be at zero. Fig.~\ref{fig:classical}(a) helps decide the position at which we should fix the angle to realize the maximum change in reflected intensity with a minimum shift in the refractive index. The reason to choose the left side of the curve is discussed earlier in the schematics section. We also provide the experimental data for a shift in angle of resonance from $0\%$ BSA (deionized water) to $10\%$ BSA solution along with the theoretical expectation for the resonance angles corresponding to different BSA concentrations in Appendix E.

Next, we lock the scanning angle at $32\degree$ to the normal at the air-prism interface (0.8 degrees left of the resonance angle) corresponding to absorption $\alpha=0.59$ absorption or the reflection coefficient $r=0.41$. We then fill the flow cell cavity with $10\%$ BSA solution that was prepared as explained in the experimental setup section. We then record the change in reflectivity as the protein binds the gold film while changing the local refractive index of the sample. As mentioned earlier the angle of resonance shifts towards the right as the refractive index increases resulting from the protein-gold binding process. Fixing the angle on the left side of the dip leads to an increase of the reflected intensity as the protein binds. This could be visualized in Fig.~\ref{fig:classical}(b), where the reflected intensity slowly increases with time and the slope of the curve gives the rate of binding. The error bars in the plot correspond to 10 consecutive readings taken at the same time stamp.

\begin{figure*}[hbt!]
\centering
\includegraphics[width=14cm]{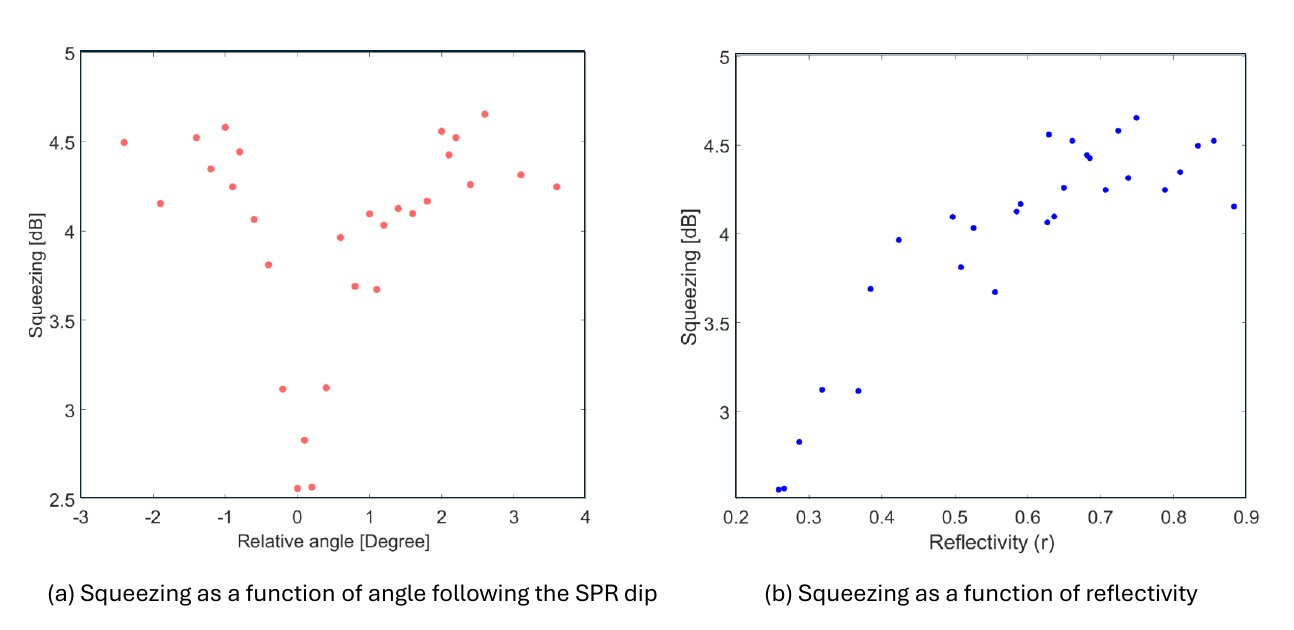}
    \caption{(a) Squeezing was measured at different scanning angles near the absorption dip. (b) This plot
shows the variation of squeezing with reflectivity, providing a clearer visualization of the expected squeezing as BSA binds to the gold. The data was derived using the reflectivity versus angle plot.}
    \label{fig:squeezing}
\end{figure*}
\textit{Squeezed light characterization}: Squeezing in dB as defined by Eq.~(\ref{Squeezing}) depends on the absorption that changes at different incident angles on the SPR sensor as shown in Fig.\ref{fig:squeezing}(a). We experimentally collected the noise power at different angles around the absorption dip and compared it with the SNL at each point. The SNL was recorded for one angle and the rest of the points were calculated using the Poisson characteristic of the coherent source $(\Delta N)^{2} = \left\langle N\right\rangle$. The coherent source is verified to be shot-noise limited around the working frequency of 2MHz as shown in Appendix D.  As shown in Fig.~\ref{fig:squeezing}(a), we get squeezing of around 4 dB at an angle of 0.8 degrees left of $\theta_{r}$ as decided using the classical SPR dip data. The squeezing as a function of reflectivity is reported in Fig.\ref{fig:squeezing}(b). 
\begin{figure*}[hbt!]
\centering
\includegraphics[width=14cm]{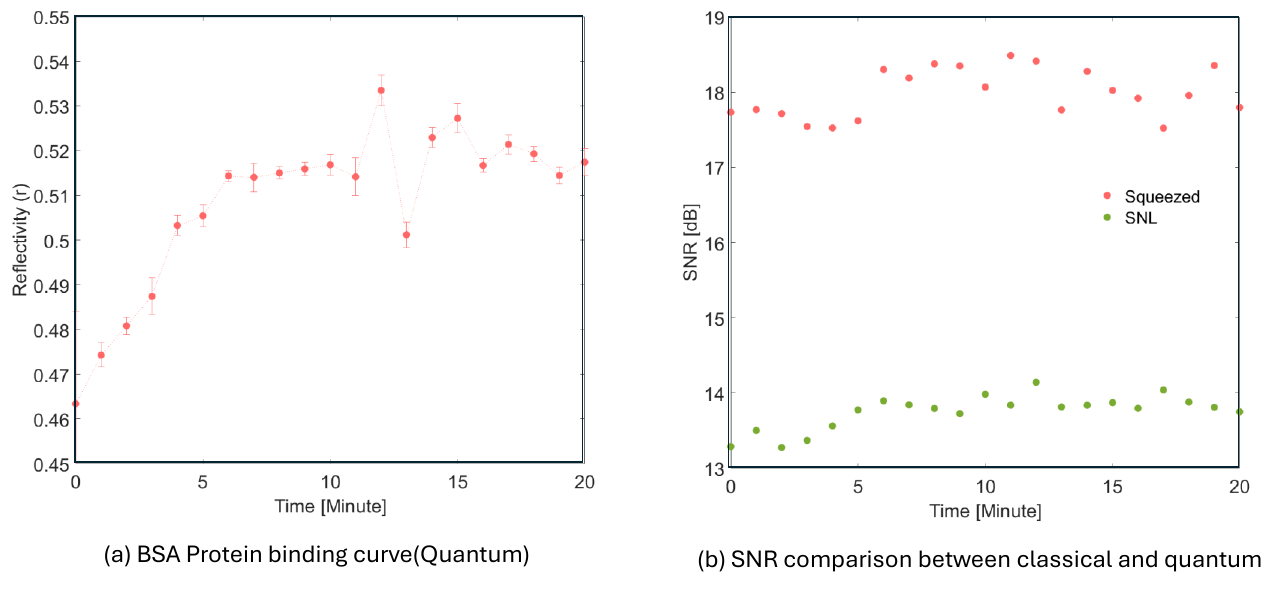}
    \caption{(a) The reflectivity of the sensor is recorded using the TMBSS beam in the probe path. The reflected intensity increases as the BSA protein binds the gold surface on the sensor.  (b) The signal-to-noise ratio calculated as explained in the detection and data processing section is plotted against time as the BSA protein binds to the gold. 4dB of quantum advantage is maintained through out the process.}
    \label{fig:Quantum}
\end{figure*}

\textit{Quantum results}: We plot the data for the protein binding process using the reflectivity of the probe beam of the TMBSS source. This plot is shown in Fig.~\ref{fig:Quantum}(a). As the protein binds, the signal increases due to the shift in SPR dip towards the right increasing the reflectivity similar to the coherent source. While taking the data, we change the conjugate manually by the same proportion as the probe. Matching the absorption on conjugate path with that of the probe path ensures that we get the maximum quantum advantage possible at that absorption level.

It is easier to see the quantum advantage in the SNR plot as a function of time reported in Fig.~\ref{fig:Quantum}(b). The quantum advantage here is defined as,
\begin{equation}
  QA = 10 \log_{10} \left(\frac{(SNR)_{\scaleto{TMBSS}{4pt}}}{(SNR)_{\scaleto{Coh}{4pt}}}\right) 
\end{equation}
The quantum advantage of the TMBSS light is realized when we take the intensity difference between the probe and the conjugate. The signal in this plot is the amplitude of the modulated peak at 2 MHz and the noise is quantified by taking the average of the side bands as shown in Fig.~\ref{fig:AOM}. We can see the quantum advantage of 4 dB despite the total absorption of $\alpha=0.74$.  The SNR for TMBSS looks flat because of a very small change in absorption. This plot demonstrates the quantum advantage of using TMBSS in applications involving biokinetic processes. 
\begin{figure}[hbt!]
\centering
\includegraphics[width=7cm]{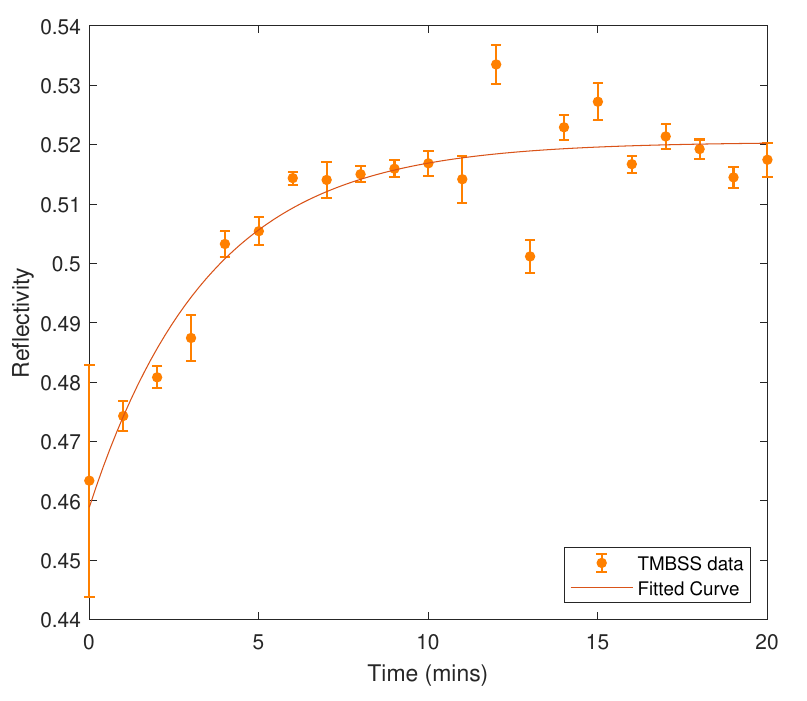}
    \caption{The nonlinear fit function was used on the probe data to extract the mean value of the parameter $k_{s}$.}
    \label{fig:datafitting}
\end{figure}

\textit{Kinetic parameter estimation sensitivity}: The interaction dynamics of the adsorption of BSA protein molecule on the gold surface and the corresponding reflectivity change can be modeled based on the framework given in \cite{xiao2019super}. The Reflected intensity denoted by R can be expressed as a function of t as follows, 
\begin{equation}
    R(t)=A+B\left(1-e^{-k_{s}t}\right)
\end{equation}
where $A$ is the initial reflected intensity at time $t=0$, $B$ is the maximum change in reflected intensity as it reaches equilibrium and $k_{s}$ is the observable rate constant in units of $s^{-1}$. The approximate mean value of the rate constant $k_{s}$ can be found out using the non-liner fit as shown in Fig.~\ref{fig:datafitting}. For our setup, we find the values as follows $A=0.4601$, $B=0.0603$ and $k_{s}=0.0046\,s^{-1}$. The statistical errors in these parameters are expected to be few percent based on the error bars shown in Fig.~\ref{fig:datafitting}. These error bars were obtained by taking 10 shots for each time t. Here we focus mainly on showing the improvement in the sensitivity of estimating the parameter $\Delta k_{s}$ when the signal is shot-noise limited.  Eq.~(7) can be written as,
\begin{equation}
    \frac{R-A}{B}=1-e^{-k_{s}t}
\end{equation}
Thus the uncertainty in R for the coherent case is given by,
\begin{equation}
    \frac{(\Delta R)_{Coh}}{B_{Coh}}= e^{-k_{s}t}(\Delta k_{s}t)_{Coh}
\end{equation}
Using Eq.(8) we can write,
\begin{equation}
    (\Delta R)_{Coh}=\left(A+B-R\right)_{Coh}(\Delta k_{s}t)_{Coh}
\end{equation}
We can write the same equation for the TMBSS case. Knowing that A,B are constants and that we are comparing the quantum and classical cases for same intensities R for a given time instant we can take the ratio to get,
\begin{equation}
    \frac{(\Delta R)_{Coh}}{(\Delta R)_{TMBSS}}=\frac{(\Delta k_{s})_{Coh}}{(\Delta k_{s})_{TMBSS}}
\end{equation}
The noise reduction in TMBSS case is attributed to the intensity difference squeezing between the probe and the conjugate whereas the mean value of signal is same for both cases. Hence the ratio in Eq.(12) could be written in terms of the Signal-to-Noise Ratio (SNR) as,
\begin{equation}
    \frac{(\Delta k_{s})_{Coh}}{(\Delta k_{s})_{TMBSS}}=\frac{(SNR)_{TMBSS}}{(SNR)_{Coh}}
\end{equation}
Using the quantum result as shown in Fig.~\ref{fig:Quantum}(b) the 4dB improvement in SNR implies the sensitivity of the estimation of $k_{s}$ is improved by $60\%$. 

In protein(Ligand)-protein(Analyte) type of interaction the rate constants are defined as  $k_{s} = k_{a} [L_{0}] + k_{d}$, where $[L_{0}]$ is the ligand concentration and $k_{a}$, $k_{d}$ are the association and dissociation rate constants respectively. As clearly explained in Eq.(1.13) in \cite{xiao2019super} analyzing the $k_{s}$ in combination with the dissociation rate analysis after the equilibration time can be used to find the binding affinity $K_{A}=\frac{k_{a}}{k_{d}}$. This parameter $K_{A}$ or its reciprocal $K_{D}$ are direct indicators of how strongly the analyte (for eg. drug protein) binds to the ligand (target protein). This is an invaluable source of information in field of drug design and discovery, and improving the sensitivity of measuring these parameters can greatly enhance drug's efficacy and safety. 
\section{Conclusion}
In summary, we have demonstrated the quantum advantage of using Two Mode Bright Squeezed State (TMBSS) in improving the sensitivity of a Surface Plasmon Resonance (SPR) based sensor to study biokinetic processes. As an example of one such process, we used the protein – Bovine Serum Albumin (BSA) binding to the gold film on the surface of the sensor. We showed 4dB of squeezing (difference between the coherent source noise and the twin beam noise) as we recorded the signal-to-noise ratio as the function of time and is maintained throughout the binding process. The quantum advantage as shown in terms of squeezing is achieved despite the total absorption of $74\%$ from the source until the final detection after the sensor. We successfully demonstrated how reduction in the state noise of the source contributes to a $60\%$ improvement in the sensitivity of estimation of the observable rate constant $k_{s}$ which is a crucial parameter in finding the affinity $K_{A}$ of a protein binding to another protein. 

Our work also reinforces the utility of TMBSS source highlighting its primary advantage: the continuous variable nature as opposed to the single photon sources that allows for a greater number of photons while still employing the quantum mechanical benefits that arise from the intensity difference squeezing. TMBSS constitutes a readily available quantum probe which is better than the coherent source because of its sub-shot noise characteristic thus allowing for lower intensities with comparable results, making it more gentle to the biological samples. The combination of TMBSS along with the sub-diffraction-limited sensing property of the SPR presents a highly promising approach in the realm of molecular and cellular biology. 

Thus, integration of a well stabilized source of TMBSS with SPR can achieve more precise measurements of protein-gold binding rate surpassing the limitations of the current shot-noise limited techniques. This improvement in the sensitivity as presented in our study can be easily extended to the kinetic analysis of the protein-protein (ligand-analyte) binding processes that are very significant in the field of biochemistry especially in drug design and discovery. Overall, this project brings together two state-of-the-art techniques and provides a most practical, quantum-enhanced SPR setup aimed at improving the sensitivity of time-dependent measurements in the study of biological processes. 

\begin{acknowledgments}
The authors are grateful for the support of the Air Force Office of Scientific Research (Award No. FA-9550-20-1-0366) and the Robert A. Welch Foundation (A-1943-20210327), National Science Foundation (Award No.2426699). The final measurements for the SNR were done in the quantum sensing lab at the University of Tennessee at Chattanooga (UTC). We wish to acknowledge the technical help provided by Charles Wallace at Texas A$\&$M University.
\end{acknowledgments}

\section*{Appendix A: Transfer-Matrix method }

\begin{figure}[H]
\centering
\includegraphics[width=7cm]{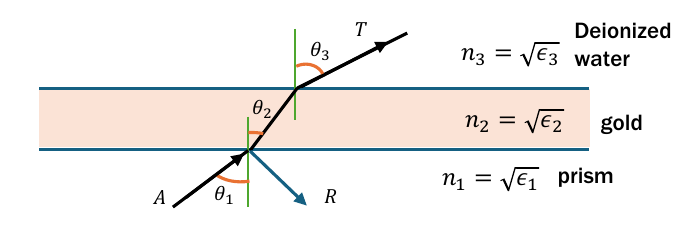}
    \caption{Propagation of EM wave through homogeneous film situated between two homogeneous media.}
    \label{fig:layered media}
\end{figure}

The Kretschmann geometry of the SPR sensor can be considered as a multilayer system where gold film with refractive index $n_2$ is situated between two semi-infinite media namely the prism with refractive of glass $n_1$ and the sample for example deionized water with refractive index of $n_3$ as shown in Fig.~\ref{fig:layered media}. The reflectance and transmission of this multilayer system can be calculated using the Transfer-Matrix method (TMM) \cite{born2013principles}. TMM is a systematic way of finding the solutions for the electromagnetic wave for appropriate boundary conditions at each interface in the stratified medium. The stratified medium is characterized by a matrix $M'$ called as the characteristic matrix of the system. The reflectance curve as a function of incident angle shown in Fig.~\ref{fig:spr}(b) is plotted using this method.
Here, the characteristic matrix $M'$ of the stratified medium evaluated at the gold-prism interface is given by,
 \begin{equation}
    M' = \begin{bmatrix}
    m'_{11} & m'_{12} \\
    m'_{21} & m'_{22}
    \end{bmatrix}=\begin{bmatrix}
     \cos \beta & -\frac{i}{q_2} \sin \beta \\
     -i q_2 \sin \beta & \cos \beta
    \end{bmatrix}\nonumber
\end{equation}
where,
\begin{equation*}
    \beta = h \sqrt{\epsilon_2 \left(\frac{\omega}{c}\right)^2 -  \epsilon_1\left(\frac{\omega}{c}\right)^2 \sin^2 \theta_1}
\end{equation*}
and
\begin{equation}
    q_l = \frac{\sqrt{\mu_l}}{\sqrt{\epsilon_l}} \cos \theta_l, \quad \text{For (TM wave)}\nonumber
\end{equation}
Here, $\epsilon_{l}$ is the dielectric permittivity of the medium $l=1,2,3$, $h$ is the thickness of the gold film and $\omega$ is the frequency of the incident field. In our calculations, the magnetic permeability $\mu_{l}$ is considered to be 1 since the magnetic field is assumed to be not interacting in this phenomenon. $\theta_l$ with $l=1,2,3$ are the angles made by the EM field with the normal for the respective mediums as shown in Fig.~\ref{fig:layered media}.

Using TMM the reflection coefficient r at the interface between gold and prism is given by the equation,
\begin{equation}
    r = \frac{R}{A} = \frac{(m'_{11} + m'_{12} q_3) q_1 - (m'_{21} + m'_{22} q_3)}{(m'_{11} + m'_{12} q_3) q_1 + (m'_{21} + m'_{22} q_3)}\nonumber
\end{equation}
where A and R are the amplitudes of the incident and the reflected electric field respectively.

The reflectivity which is the $|r|^{2}$ is then plotted against $\theta_{1}$. This quantity reaches its minimum at the resonance angle where the energy from the incident electric field is mostly absorbed by the excited surface plasmon polaritons. While calculating the SPR angle we also consider the Fresnel coefficients at the prism-air interface at which the EM field enters and exits the prism. Thus we see the reflectivity in Fig.~\ref{fig:spr}(b) for the deionized water reaches its minimum at the angle of 66.73 degrees and shifts towards right after the BSA binds the gold. 

\section*{Appendix B: Four-Wave Mixing process} 
The Four Wave Mixing process crucial in generating the TMBSS takes place inside a 2.5 cm long cylindrical vapor cell containing Rb-85 atomic ensemble. The atomic vapor serves as a non-linear medium where the $D_{1}$ line corresponding to the $5S_{1/2} \rightarrow 5P_{1/2}$ transition including the hyperfine structure of the Rb85 atom constitutes a double-$\Lambda$ energy level configuration as shown in Fig.~\ref{fig:exp setup}(b).

As shown in the upper half portion of Fig.~\ref{fig:exp setup}(a), we use a continuous wave laser of $\lambda=795nm$ which is split into two beams. One of them is used as a strong pump of $450mW$ to drive the FWM process and the other is used as a seed probe which is a cross-polarized weak beam of $150 \mu W$ to enhance the FWM process. As explained in Fig.~\ref{fig:exp setup}(b), this process requires tuning of both the frequencies where the phase matching condition given by $2\omega_{P} = \omega_{Pr}+\omega_{Cj}$ along with the momentum conservation condition is satisfied  the Rubidium cell. The pump frequency $\omega_{p}$ is $\Delta$ detuned from the Doppler broadened excited state which drives the atom from f=2 to the f=3 ground state via stokes scattered photon of frequency $\omega_{Pr}$. Similarly, when another pump photon drives the atom back into the f=2 ground state the anti-stokes scattered photon is generated with frequency $\omega_{Cj}$. This process is enhanced by the usage of the seed probe with the frequency $\omega_{Pr} = \omega_{p}-(\nu_{HF}+\delta)$ where $\nu_{HF}$ is the hyperfine splitting of the electronic ground sate and $\delta$ is the two photon detuning as shown in the energy level diagram.

Single photon detuning $\Delta$ is achieved by locking the main laser by approximately 1 GHz away from the absorption dip recorded using a reference Rb cell. The Two-photon detuning $\delta$ on the seed probe frequency is achieved by using a double pass AOM of 1.52 GHz. $\delta$ ensures that the time lag between probe and conjugate is very small. The momentum conservation is taken care of by carefully aligning the pump beam and the weak probe beam using appropriate mirror arrangements. Another important parameter necessary for FWM process is the temperature of the cell. The cell window is maintained at around 140\degree C to prevent the Rb atoms from settling on the window whereas the core temperature is maintained around 90\degree C as required for the process of FWM. This leads to the simultaneously generated, equally polarized probe and the conjugate photons that share quantum correlations and are referred to as being in the Two-Mode Bright Squeezed State.

\section*{Appendix C: Sensor assembly}
We prepared the sensor using the BK7 (n=1.51) right angle prism from Thorlabs, a microscope glass slide with 50 nm gold layer (SKU: AU.0500.ALSI) bought from Platypus technologies, and the customized flow cell designed and fabricated using on campus 3D printing facility.  As shown in Fig.~\ref{fig:exp setup}(c), the sample delivery system needs to work such that the proteins get in contact with the gold without leaking from the flow cell. We use O-ring to avoid any leakage from the flow cell as it is tightly clamped against the microscope slide and the prism. The input and the output ports of the flow cell are connected to a peristaltic pump to control the flow of the sample.
Another challenge during the clamping process is the formation of air bubbles between the glass slide and the prism. We use immersion oil (56822-50ml Sigma Aldrich, n=1.5) to avoid any air bubbles. We chose the refractive index of the glass portion of the microscope slide same as the refractive index of prism as well as the immersion oil such that there is no refraction at the gold-dielectric interface.
\begin{figure}[htbp]
\centering
\includegraphics[width=8cm]{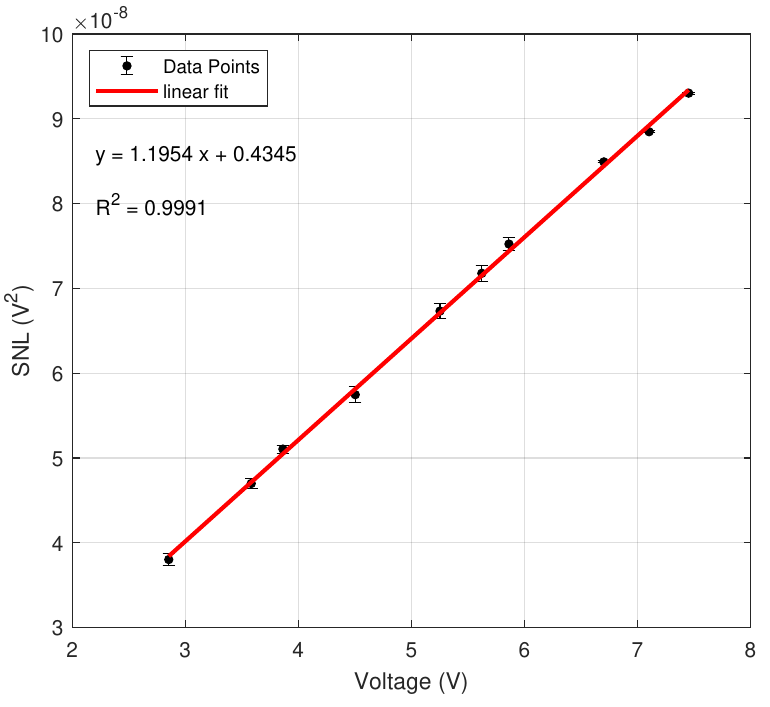}
    \caption{Shot-noise limit verification by plotting the variance ($V^2$) of the signal associated with the coherent source against the total intensity ($V$) of the source entering the detectors A and B.}
    \label{fig:SNL verification}
\end{figure}

\section*{Appendix D: Shot Noise Limitation} 
We verified the coherent source for its shot-noise limit. The two coherent beams were directed in the two detectors A and B replicating the arrangement for TMBSS source. The verification was done using the oscilloscope for voltage readings and the spectrum analyzer was used to record the noise power around the working frequency of 2MHz. As shown in Fig.~\ref{fig:SNL verification}, units for y-axis were converted from the dBm units as displayed on the spectrum analyzer to $V^2$ and plotted against the total voltage $(V)$. The difference in the scale of the units can be ascribed to using two different instruments as oscilloscope with $1 M\ohm$ impedance while the spectrum analyzer with $50 \ohm$. The linear relation between the mean of the signal $(V)$ and its variance $(V^2)$ confirms that our study was done in a shot noise limited domain.

\section*{Appendix E: SPR calibration curve}

\begin{figure}[htbp]
\centering
\includegraphics[width=7cm]{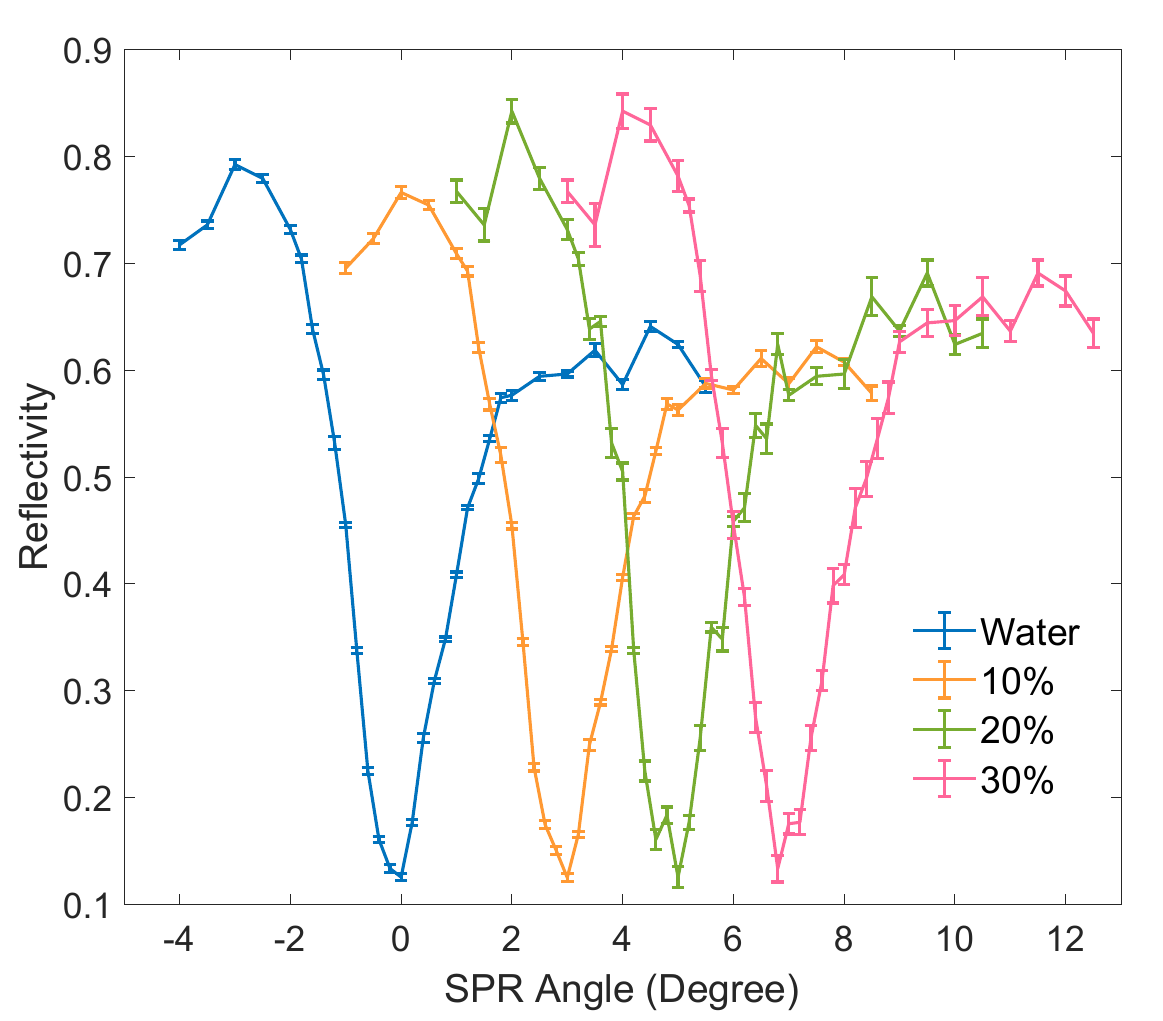}
    \caption{Reflectivity of the sensor is plotted against the  external scanning angle demonstrating the shift in the resonance angle from $0\%$ BSA (deionized water) to $10\%, 20\%, 30\%$ BSA solution.}
    \label{fig:Shift}
\end{figure}

\begin{figure}[htbp]
\centering
\includegraphics[width=7cm]{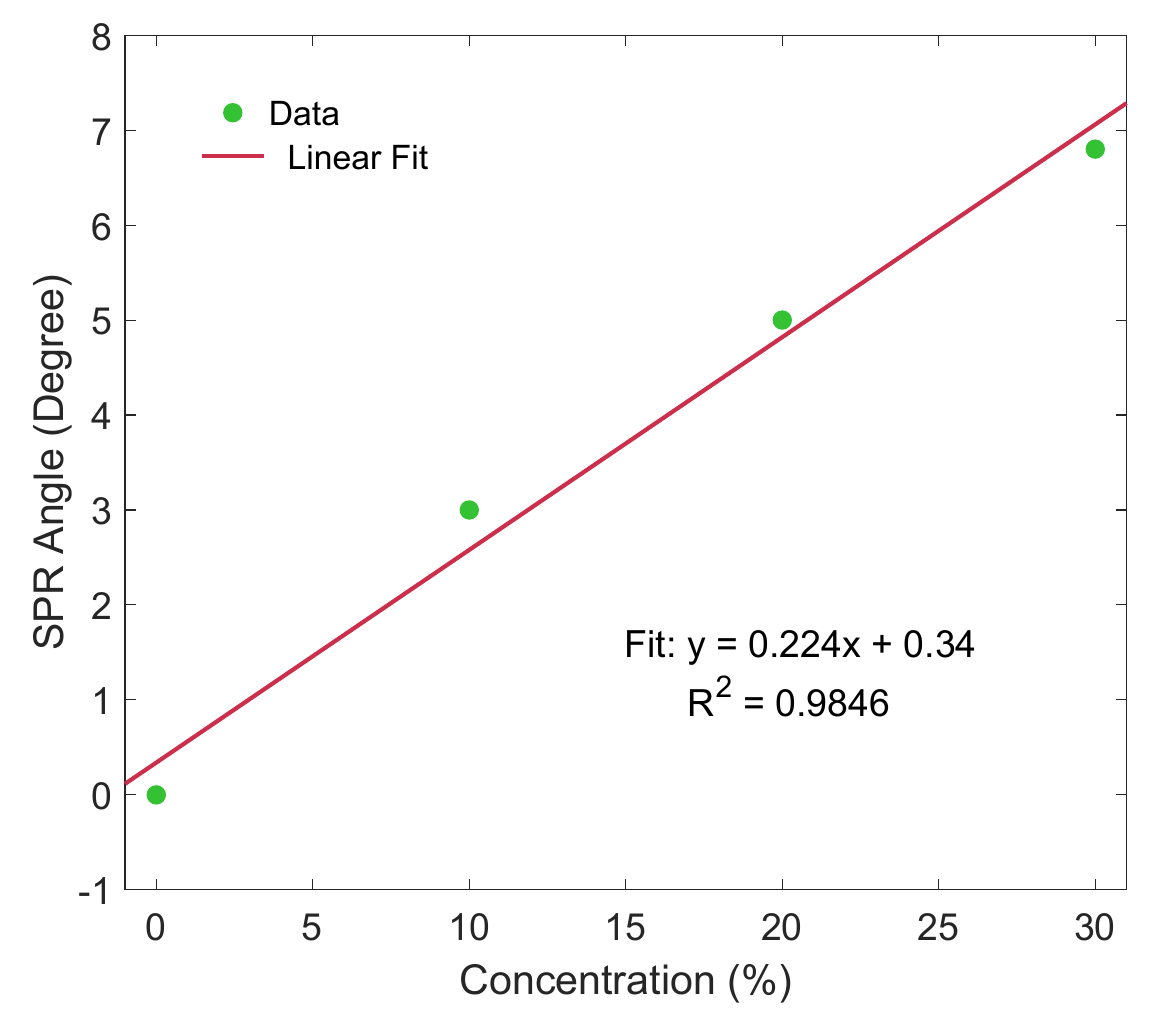}
    \caption{SPR calibration curve : The angle of resonance was plotted for different concentrations of the BSA solution. }
    \label{fig:calibration curve exp}
\end{figure} 

We perform the SPR Calibration under the exact instrumental conditions used for protein adsorption measurements by recording the reflectivity of the sensor against the external scanning angle for $0\%$ BSA (deionized water) and $10\%, 20\%, 30\% $ BSA solutions as shown in Fig.~\ref{fig:Shift} using the coherent source. The scale on X-axis is calibrated such that the resonance angle for deionized water is at zero. 

We then plot the resonance angle against each concentration as shown in Fig.~\ref{fig:calibration curve exp}. The plot shows the linearity as expected for a calibration curve. We compare this analysis with the theoretical expectations derived using the information provided in \cite{barer1954refractive} about the refractive indices of various concentrations of the BSA solution. We map these refractive indices to the angle of resonance calculated using the Transfer Matrix Method (TMM) as explained in appendix A and the schematics section in this paper. We expect to see an approximate 2-degree shift as the percentage concentration increases by $10\%$ as confirmed in Fig.~\ref{fig:calibration curve exp}.




\bibliography{main}

\end{document}